\documentclass[aip,jcp,reprint,amsmath,amssymb,nobalancelastpage]{revtex4-2}
\usepackage{bm}
\usepackage{graphicx}
\usepackage[suffix=]{epstopdf}
\usepackage{natmove}

\newcommand{\la}{\left\langle}
\newcommand{\ra}{\right\rangle}

\newcommand{\vv}[1]{\mathbf{#1}}
\renewcommand{\d}[1]{\ensuremath{\operatorname{d}\!{#1}}}

\begin{document}
\title{Diffusion and sedimentation in colloidal suspensions using multiparticle collision dynamics with a discrete particle model}

\author{Yashraj M. Wani}
\affiliation{Institute of Physics, Johannes Gutenberg University Mainz, Staudingerweg 7, 55128 Mainz, Germany}

\author{Penelope Grace Kovakas}
\affiliation{Department of Chemical Engineering, Auburn University, Auburn, AL 36849, USA}

\author{Arash Nikoubashman}
\email{anikouba@uni-mainz.de}
\affiliation{Institute of Physics, Johannes Gutenberg University Mainz, Staudingerweg 7, 55128 Mainz, Germany}

\author{Michael P. Howard}
\email{mphoward@auburn.edu}
\affiliation{Department of Chemical Engineering, Auburn University, Auburn, AL 36849, USA}

\begin{abstract}
We study self-diffusion and sedimentation in colloidal suspensions of nearly-hard spheres using the multiparticle collision dynamics simulation method for the solvent with a discrete mesh model for the colloidal particles (MD+MPCD). We cover colloid volume fractions from $0.01$ to $0.40$ and compare the MD+MPCD simulations to Brownian dynamics simulations with free-draining hydrodynamics (BD) as well as pairwise far-field hydrodynamics described using the Rotne--Prager--Yamakawa mobility tensor (BD+RPY). The dynamics in MD+MPCD suggest that the colloidal particles are only partially coupled to the solvent at short times. However, the long-time self-diffusion coefficient in MD+MPCD is comparable to that in BD and BD+RPY, and the sedimentation coefficient in MD+MPCD is in good agreement with that in BD+RPY, suggesting that MD+MPCD gives a reasonable description of the hydrodynamic interactions in colloidal suspensions. The discrete-particle MD+MPCD approach is convenient and readily extended to more complex shapes, and we determine the long-time self-diffusion coefficient in suspensions of nearly-hard cubes to demonstrate its generality.
\end{abstract}

\maketitle

\section{Introduction}
The dynamics of nanometer- or micrometer-sized colloidal particles (``colloids'') suspended in a solvent control many processes, ranging from the nanoscale assembly of functional materials \cite{boles:chemrev:2016} and transport in biological systems \cite{sear:prl:2019} to industrial-scale processes such as enhanced oil recovery \cite{shamsijazeyi:2014}. The motion of the colloids is dictated by not only direct interactions between the suspended particles, such as dispersion forces or electrostatics, but also fluctuating hydrodynamic forces exerted by the solvent \cite{russel:1989}. The many-body hydrodynamic interactions in colloidal suspensions are typically described theoretically using a low-Reynolds-number treatment of the solvent \cite{happel:1983, kim:2005}. These equations are challenging to solve for all but a handful of special theoretical cases, so numerical approaches are frequently required. Here, we restrict our discussion to particle-based computer simulation techniques that are well-suited for describing suspensions with thermal fluctuations and complex chemistries.

Particle-based computer simulations that fully resolve both the suspended colloids and solvent (e.g., atomistic molecular dynamics) are prohibitively expensive for reaching the long time scales and large length scales that are practically relevant, necessitating the use of coarse-grained models.\cite{howard:coce:2019} In these models, the colloids are resolved explicitly while the thermodynamic and hydrodynamic effects of the solvent are captured either implicitly or using a simplified description. Brownian dynamics (BD) is a classic implicit-solvent technique \cite{allen:2017} that, in its simplest form, neglects hydrodynamic forces other than single-colloid drag (the free-draining approximation). Hydrodynamic forces between colloids can be introduced to BD by an appropriate hydrodynamic tensor \cite{ermak:jcp:1978}, as in the Rotne--Prager--Yamakawa \cite{rotne:jcp:1969, yamakawa:jcp:1970} (BD+RPY) or Stokesian dynamics approaches \cite{brady:jfm:1988, brady:1988}. These techniques are powerful and computationally efficient because they do not need to capture details of the solvent; however, the correct hydrodynamic tensor must be known, which can be challenging for complex particle shapes or bounded geometries.

An alternative family of ``mesoscale'' simulation methods, which includes multiparticle collision dynamics (MPCD),\cite{malevanets:jcp:1999, gompper:adv:2009, howard:coce:2019} dissipative particle dynamics \cite{hoogerbrugge:epl:1992, groot:1997}, and the lattice Boltzmann method \cite{mcnamara:prl:1988, ladd:jsp:2001}, retains an explicit solvent but uses a simplified model with similar physical properties as the real solvent. The colloids are coupled to the solvent so that hydrodynamic forces between them emerge naturally within the model. Hence, these approaches are generally more compatible with complex particles and geometries; however, there are still active questions about how the solvent properties and coupling scheme affect the hydrodynamic forces, and in particular, whether they reproduce low-Reynolds-number results in appropriate limits \cite{marson:sm:2018, fiore:prf:2018, shakeri:pfl:2018}. The focus of this article is on the MPCD method, which is the newest of the three mesoscale approaches listed above. In MPCD, the solvent is modeled as point particles that interact with each other through stochastic collisions in spatially localized cells \cite{malevanets:jcp:1999}. Depending on the collision scheme, the MPCD solvent has properties of a gas-like or liquid-like (typically Newtonian) fluid \cite{ripoll:pre:2005}.

Numerous approaches have been proposed for coupling colloids to the MPCD solvent. Several studies modeled the colloids as smooth spheres that interact with the solvent through a short-ranged repulsive pair potential;\cite{malevanets:jcp:2000, padding:pre:2006, batot:pre:2013, dahirel:pre:2018} however, this coupling can only give slip boundary conditions on the colloids. Inoue et al.~proposed an alternative approach where the solvent is stochastically scattered from the colloids,\cite{inoue:jstat:2002} transferring both translational and rotational momentum  to produce no-slip boundary conditions. The resulting dynamics of a single colloid were in good agreement with theoretical predictions \cite{padding:jpcm:2005} while the colloid long-time self-diffusion coefficients in suspensions were slightly larger than in reference simulations using fast lubrication dynamics \cite{bolintineanu:cpm:2014}. This coupling scheme has been used to study many processes including sedimenting \cite{wysocki:faraday:2010} and sheared \cite{hecht:pre:2005, nikoubashman:sm:2015} suspensions. We note, however, that the stochastic reflection procedure approximates the colloid surface as an infinite plane, which limits its application to particles with low curvature and when the mean free path of the solvent particles is small compared to the colloid diameter;\cite{hecht:pre:2005} this assumption can be relaxed by reflecting the solvent from the surface as an elastic collision \cite{yang:sm:2014}.

All of these reflection-based coupling schemes suffer from a few practical drawbacks. First, reflection may introduce spurious depletion forces between colloids that must be carefully handled \cite{hecht:pre:2005, padding:pre:2006}. Second, the presence of impenetrable solid surfaces inevitably leads to partially filled collision cells; this alters the local solvent properties near the surfaces and requires the insertion of ``virtual'' solvent particles inside the moving colloids \cite{lamura:epl:2001, bolintineanu:pre:2012, whitmer:jpcm:2010}. Last, performing the reflection requires pairwise collision detection between the solvent and colloids that is computationally expensive due to the large number of solvent particles. These issues are exacerbated in dense suspensions.

To circumvent these problems, Poblete et al.~recently proposed a discrete particle model \cite{poblete:pre:2014} that represents a colloid as a shell of point particles (``vertex particles'') connected by elastic springs (Fig.~\ref{fig:model}).\cite{lobaskin:njp:2004, fischer:jcp:2015, swan:pfl:2016} Unlike in reflection-based coupling schemes, the colloid is fully penetrable to the solvent, and the vertex particles couple to the solvent only during the stochastic collision. In addition to the benefits of addressing physical and computational challenges associated with reflection-based coupling, this model is compatible with arbitrary colloid shapes such as spherocylinders \cite{kobayashi:sm:2020} and with complex surface patterns \cite{kobayashi:sm:2020, kobayashi:lng:2020}. The hydrodynamic properties of a {\it single} such colloid were studied for various degrees of discretization, finding good agreement with low-Reynolds-number theoretical predictions when one vertex particle occupied approximately one collision cell \cite{poblete:pre:2014}. To our knowledge, however, the properties of suspensions represented using this colloid model in MPCD have not been studied in detail. Characterizing the transport properties of colloidal suspensions is not only of fundamental interest in its own right but also important for reliably applying similar models to more complex soft-matter systems.

Here, we perform bulk simulations of a colloidal suspension of nearly-hard spheres using MPCD with a discrete particle model. We characterize the short-time and long-time self-diffusive motion from equilibrium simulations as well as the sedimentation coefficients from nonequilibrium simulations. We compare these simulations to BD and BD+RPY simulations as well as low-Reynolds-number theoretical predictions as functions of volume fraction, finding good agreement for some quantities but discrepancies in others. To demonstrate the generality of the discrete particle model, we also simulate a colloidal suspension of nearly-hard cubes and report the long-time self-diffusion coefficient as a function of volume fraction. Overall, we find that MPCD with a discrete particle model is convenient to implement and gives a good description of colloidal suspension dynamics.

\section{Models}
\label{sec:models}

\subsection{Multiparticle collision dynamics}
\label{sec:models:solvent}
The MPCD solvent consists of point particles, each having a position $\vv{r}_i$, a velocity $\vv{v}_i$, and the same mass $m$, whose motion is governed by alternating streaming and collision steps \cite{gompper:adv:2009, howard:coce:2019}. During the streaming step, the solvent particles move according to Newton's equations of motion
\begin{align}
    \frac{\d{\vv{r}_i}}{\d{t}} &= \vv{v}_i \nonumber \\
    m_i \frac{\d{\vv{v}_i}}{\d{t}} &= \vv{F}_i ,
    \label{eq:MPCDStream}
\end{align}
where $m_i$ is the mass of particle $i$ and $\vv{F}_i$ is the force acting on it. In the case of a constant (or zero) force acting on each particle, these equations may be analytically integrated to give the equations of ballistic motion, while more complex forces require a numerical integration scheme \cite{allen:2017}.

A stochastic multiparticle collision is performed at equal time intervals between streaming steps. During this collision step, each particle is assigned to a cubic cell of edge length $\ell$ to undergo a spatially localized momentum exchange. To ensure Galilean invariance \cite{ihle:pre:2001}, the collision cells are shifted along each Cartesian axis by a random amount drawn uniformly on $[-\ell/2, +\ell/2]$. In this work, we apply the stochastic rotation dynamics (SRD) collision scheme without angular momentum conservation \cite{malevanets:jcp:1999}, where the particle velocities are updated during the collision by
\begin{equation}
    \vv{v}_i \gets \vv{u}_j + \bm{\Omega}_j \cdot \left(\vv{v}_i - \vv{u}_j\right) ,
    \label{eq:MPCDCollision}
\end{equation}
where $\mathbf{u}_j$ is the center-of-mass velocity and $\bm{\Omega}_j$ is the rotation operator of the cell $j$ that contains particle $i$. The rotation operator is chosen to have a fixed rotation angle but a random rotation axis for each cell. During the collision, we apply a cell-level Maxwellian thermostat to maintain a constant temperature $T$ \cite{huang:jcp:2010}. Both the streaming and collision steps conserve linear momentum and approximately reproduce hydrodynamic interactions in the solvent down to the cell size \cite{huang:pre:2012, dahirel:pre:2018}.

The natural system of units describing the MPCD solvent is defined by the mass $m$ of a solvent particle, the length $\ell$ of a collision-cell edge, and the energy scale $k_{\rm B} T$; the resulting unit of time is $\tau = \sqrt{m \ell^2 \beta}$ with $\beta = 1/(k_{\rm B} T)$. We adopted standard SRD parameters---collision time $0.1\,\tau$, average solvent number density $\rho_0 = 5\,\ell^{-3}$, and rotation angle $130^\circ$---resulting in a liquid-like Newtonian fluid with dynamic viscosity $\eta_0 = 3.95 \pm 0.01\,k_{\rm B}T\tau/\ell^3$ and kinematic viscosity $\nu_0 = \eta_0/(\rho_0 m) = 0.79\,\ell^2/\tau$.\cite{statt:prf:2019}

\subsection{Discrete particle model for colloids}
\begin{figure}
    \centering
    \includegraphics{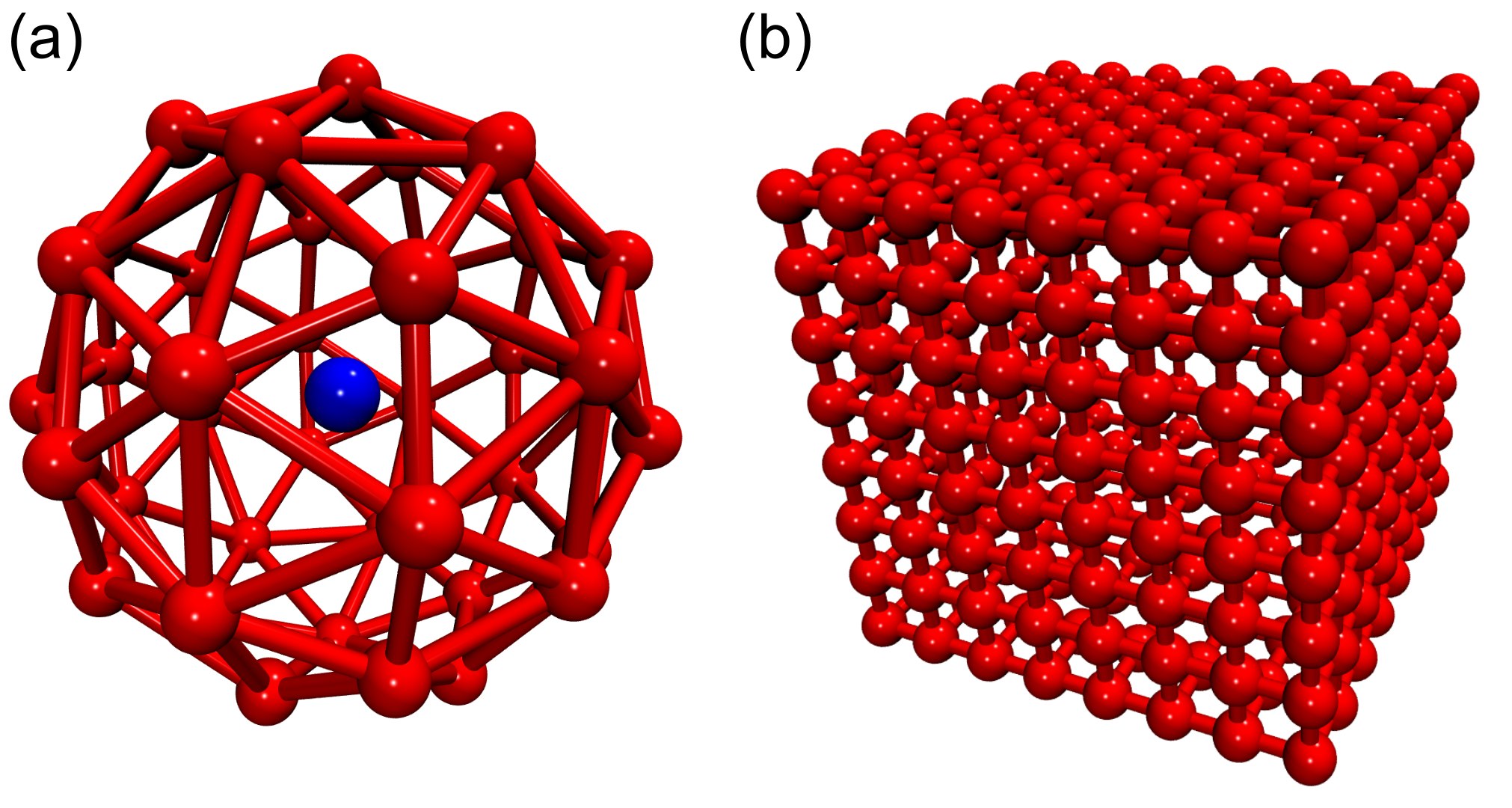}
    \caption{Discrete particle model for (a) spherical colloid with radius $a=3\,\ell$ and (b) cubic colloid with diameter $d=6\,\ell$. The $N_{\rm v}$ vertex particles and bonds between nearest neighbors are shown in red. The central particle for the spherical colloid is shown in blue; particle sizes have been reduced for clarity. Rendered using Visual Molecular Dynamics (version 1.9.3).\cite{vmd}}
    \label{fig:model}
\end{figure}

The spherical colloids were coupled to the MPCD solvent using a discrete particle model [Fig.~\ref{fig:model}(a)] \cite{poblete:pre:2014}. Each colloid was represented by $N_{\rm v}$ vertex particles on the surface of a sphere of radius $a$. An additional particle was placed at the center of the sphere and used to compute excluded-volume interactions between colloids (see below). The positions of the vertices were generated by recursively subdividing the faces of a regular icosahedron into equilateral triangles, then radially scaling the vertices to a distance $a$ from the center of the sphere. Note that the scaling procedure distorts the triangular faces so that the separations between neighboring vertices may no longer be equal. The resulting shape is sometimes referred to as an icosphere.

To maintain the surface-mesh shape, the vertex particles were connected with their nearest neighbors and with the central particle using a harmonic potential,
\begin{equation}
    \beta u_{\rm b}(r) = \frac{k}{2}(r-r_{\rm b})^2 ,
    \label{eq:Ub}
\end{equation}
where $r$ and $r_{\rm b}$ are the actual and desired distances between bonded particles, respectively. We set the spring constant to $k = 5000\,\ell^{-2}$ to ensure the (nearly) rigid shape of the colloid \cite{poblete:pre:2014}. In this work, we used sphere radius $a = 3\,\ell$ and subdivided the icosahedron once, resulting in $N_{\rm v}=42$ vertex particles and bond lengths $r_{\rm b} = 1.64\,\ell$, $1.85\,\ell$, and $3\,\ell$. We set the mass of each vertex particle and the central particle equal to the average mass of solvent in a collision cell ($m_{\rm v} = 5\,m$), giving total colloid mass $M = (N_{\rm v}+1) m_{\rm v} = 215\,m$.

To represent hard-sphere excluded-volume interactions between colloids, the central particles interacted through the core-shifted Weeks--Chandler--Andersen potential \cite{weeks:jcp:1971},
\begin{equation}
    \beta u(r) = \begin{cases}
        \displaystyle 4 \left[\left(\frac{\sigma}{r-\Delta}\right)^{12} - \left(\frac{\sigma}{r-\Delta}\right)^6 \right] + 1,& r \le r^* \\
        0,& {\rm otherwise}
    \end{cases} ,
    \label{eq:wca}
\end{equation}
where $r$ is the distance between two central particles, and we used $\sigma = 1.0\,\ell$, $\Delta = 5.0 \,\ell$, and $r^* = \Delta + 2^{1/6} \sigma \approx 6.12\,\ell$. This potential is purely repulsive and gives an effective hard-sphere diameter of $6.02\,\ell$ upon mapping the second virial coefficient, in good agreement with the nominal diameter $d = 6\,\ell$.

The vertex particles (but not the central particle) were coupled to the MPCD solvent through the collision step \cite{poblete:pre:2014} [Eq.~\eqref{eq:MPCDCollision}] with appropriate treatment of the different particle masses when computing the center-of-mass velocity of a cell and applying the thermostat. Between collisions, the positions and velocities of the vertex particles and the central particle were evolved according to Newton's equations of motion [Eq.~\eqref{eq:MPCDStream}] using a velocity Verlet integration scheme with timestep $0.005\,\tau$, as in traditional molecular dynamics (MD) simulations \cite{allen:2017}. Accordingly, we will refer to the discrete particle model coupled to the MPCD solvent as MD+MPCD in the rest of this article.

\subsection{Brownian dynamics}
\label{sec:models:BD}
To facilitate assessment of the MD+MPCD model, we also performed Brownian dynamics simulations using both free-draining hydrodynamics (BD) and pairwise far-field hydrodynamics (BD+RPY) \cite{allen:2017, ermak:jcp:1978, rotne:jcp:1969, yamakawa:jcp:1970}. These models are popular and have well-defined hydrodynamic interactions, and so serve as useful reference points for understanding the solvent-mediated interactions that emerge in the MD+MPCD model. A single colloid at infinite dilution in a Newtonian fluid diffuses according to the Stokes--Einstein relationship, $D_0 = k_{\rm B} T/\gamma_0$, at times much longer than the inertial (Brownian) timescale $\tau_{\rm B} = M/\gamma_0$. For spheres with no-slip boundary conditions, the friction coefficient $\gamma_0 = 6\pi\eta_0 a$, so $D_0 = 4.48 \times 10^{-3}\,\ell^2/\tau$ and $\tau_{\rm B} = 0.96\,\tau$ for the MPCD solvent  (Sec.~\ref{sec:models:solvent}). This inertial time is much less than the time required for a colloid to diffuse its own radius $\tau_0 = a^2/D_0 \approx 2.0\times 10^3\,\tau$, justifying the neglect of colloid inertia that is implicit to Brownian dynamics simulations.

We represented each colloid using only the central particle, as vertex particles are not required for Brownian dynamics simulations of bulk suspensions of spheres, using the same colloid interactions given by Eq.~\eqref{eq:wca}. Considering only the translation of the colloids in the low-Reynolds-number limit, the vector of positions of all colloids $\mathcal{R} = (\vv{r}_1,\vv{r}_2,\dots)$ changes by $\Delta \mathcal{R}$ over a time $\Delta t$ according to \cite{ermak:jcp:1978}
\begin{equation}
    \Delta \mathcal{R} = \left[\mathcal{M} \cdot \mathcal{F} + k_{\rm B} T (\nabla \cdot \mathcal{M})\right] \Delta t + \Delta\mathcal{W},
\label{eq:BD}
\end{equation}
where $\mathcal{M}$ is the mobility tensor coupling the translation of each colloid to the vector of forces on all of the colloids $\mathcal{F} = (\vv{F}_1, \vv{F}_2, \dots)$, and $\Delta\mathcal{W}$ is the vector of random displacements of all colloids having a Gaussian distribution with zero mean and covariance $\la \Delta\mathcal{W} \Delta\mathcal{W} \ra = 2 k_{\rm B} T \mathcal{M} \Delta t$.  The divergence of $\mathcal{M}$ is taken with respect to $\mathcal{R}$ \cite{ermak:jcp:1978, wajnryb:physa:2004}.

The hydrodynamic interactions that are present in the simulations are determined by the choice of $\mathcal{M}$. Stokesian dynamics includes both near-field (lubrication) and far-field coupling in $\mathcal{M}$ and can also include torques and stresslets \cite{brady:1988}, but has a corresponding large computational cost. Here, we use the simpler but reasonably accurate approximation that the entries of $\mathcal{M}$ are given pairwise,
\begin{equation}
    \mathcal{M} = \begin{pmatrix}
    \vv{M}^{(11)} & \vv{M}^{(12)} & ... \\
    \vv{M}^{(21)} & \vv{M}^{(22)} & ... \\
    \vdots & \vdots & \ddots
    \end{pmatrix} ,
\end{equation}
where $\vv{M}^{(ij)}$ couples the velocity of colloid $i$ to the force on colloid $j$ and depends on only $\vv{r}_i$ and $\vv{r}_j$. With free-draining hydrodynamics (BD), $\vv{M}^{(ii)} = \gamma_0^{-1} \vv{I}$ and $\vv{M}^{(ij)} = \vv{0}$ for $i \ne j$ so that each colloid experiences only its bare particle drag and no hydrodynamic interactions with the other colloids. ($\vv{I}$ is the identity tensor.) With pairwise far-field hydrodynamics (BD+RPY), we use a periodic formulation \cite{hasimoto:jfm:1959} of the free-space RPY tensor \cite{rotne:jcp:1969, yamakawa:jcp:1970}, $\vv{M}^{(ii)} = \gamma_0^{-1} \vv{I}$ and
\begin{equation}
    \vv{M}^{(ij)} = \gamma_0^{-1} \begin{cases}
    \displaystyle \left(\frac{3a}{4r}+\frac{a^3}{2r^3}\right)\vv{I} + \left(\frac{3a}{4r}-\frac{3a^3}{2r^3}\right)\hat{\vv{r}}\hat{\vv{r}},& r > d \\
    \displaystyle \left(1-\frac{9r}{32 a}\right)\vv{I} + \frac{3r}{32a}\hat{\vv{r}}\hat{\vv{r}},& r \le d
\end{cases}
    \label{eq:RPY}
\end{equation}
for $i \ne j$, where $\hat{\vv{r}} = \vv{r}/r$ is the unit vector from $j$ to $i$ ($\vv{r} = \vv{r}_i-\vv{r}_j$ and $r = |\vv{r}|$). The periodic formulation of the free-space tensor is \cite{fiore:jcp:2017}
\begin{equation}
\vv{M}^{(ij)} = \frac{1}{\eta_0 V} \sum_{\vv{k}\ne\vv{0}} e^{i\vv{k}\cdot(\vv{r}_i-\vv{r}_j)} \frac{1}{k^2}\left(\frac{\sin ka}{ka}\right)^2 (\vv{I}-\hat{\vv{k}}\hat{\vv{k}}),
\label{eq:RPYkspace}
\end{equation}
where $V$ is the volume of the simulation box, $\vv{k}$ is a wavevector commensurate with the simulation box, $k = |\vv{k}|$, and $\vv{\hat{k}} = \vv{k}/k$. The sum over wavevectors must be treated carefully when applying Eq.~\eqref{eq:BD}, so we used the efficient positively split Ewald technique of Fiore et al.~with splitting parameter $0.5$ and relative error tolerance $10^{-3}$ \cite{fiore:jcp:2017}. The mobility tensor $\mathcal{M}$ for both the BD model and the BD+RPY model is divergence-free ($\nabla \cdot \mathcal{M} = \vv{0}$) \cite{ermak:jcp:1978, wajnryb:physa:2004}, so this term can be neglected in Eq.~\eqref{eq:BD}. We used timestep $\Delta t = 10^{-5}\,\tau_0 \approx 0.02\,\tau$ in both our BD and BD+RPY simulations for numerical stability.

The natural system of units describing the colloids in Brownian dynamics is defined by the colloid radius $a$, the energy scale $k_{\rm B} T$, and the diffusion time $\tau_{\rm 0}$. As this system of units is also useful and general for describing colloidal suspensions, we will use these units rather than the MPCD units of Sec.~\ref{sec:models:solvent} to present most of our results. It is straightforward to convert between the two if desired.

\subsection{Langevin dynamics}
In addition to MPCD and BD, we also used Langevin dynamics (LD) simulations with free-draining hydrodynamics to initially equilibrate the colloid configurations and to perform selected comparisons (Sec.~\ref{sec:D:short}). The standard LD equations of motion for a particle $i$ are \cite{allen:2017}
\begin{align}
\frac{\d{\vv{r}_i}}{\d{t}} &= \vv{v}_i \nonumber \\
m_i \frac{\d{\vv{v}_i}}{\d{t}} &= \vv{F}_i - \gamma_i \vv{v}_i + \vv{F}_{{\rm R},i} ,
\end{align}
where $\gamma_i$ is the particle drag coefficient and $\vv{F}_{{\rm R},i}$ is a randomly fluctuating force having a Gaussian distribution with zero mean and covariance $\la \vv{F}_{{\rm R},i}(t) \vv{F}_{{\rm R},i}(t') \ra = 2 k_{\rm B} T \gamma_i \delta(t-t') \vv{I}$. For colloids represented by single spheres, these equations apply straightforwardly with $m_i = M$ and $\gamma_i = \gamma_0$ (BD is the overdamped limit of LD). However, they may also be applied to the discrete particle model using $m_i = m_{\rm v}$ and a desired $\gamma_i$. Taking $\gamma_i = \gamma_0/(N_{\rm v}+1)$ results in the same diffusivity for an isolated colloid, but other values can be chosen, e.g., to accelerate equilibration.

\subsection{Simulation details}
We simulated bulk suspensions containing $N$ spheres in a three-dimensional cubic simulation box of edge length $L$ with periodic boundary conditions. For most simulations, we fixed the box size at $L = 120\,\ell = 40\,a$ and varied the number of colloids to obtain a desired volume fraction $\phi = N v/ L^3$, where $v$ is the volume of a single colloid ($v = 4 \pi a^3/3$ for spherical colloids). We varied $\phi$ in the range $0.01 \le \phi \le 0.40$, giving between 153 and 6112 colloids. We also varied $L$ in some simulations to test for finite-size effects (Sec.~\ref{sec:D}). Equilibrium configurations were prepared by initially placing the discrete-particle colloids in the box without overlap, then simulating for $2.5 \times 10^4\,\tau$ using Langevin dynamics with $\gamma_i = 1.0\,m/\tau$. We saved 8 equilibrated configurations that we subsequently used to initialize production simulations for the various models. The simulations were performed using HOOMD-blue (version 2.9.6) \cite{howard:cpc:2018, anderson:cms:2020} extended with azplugins (version 0.10.2) \cite{azplugins} and a public version of the positively split Ewald method \cite{fiore:jcp:2017, pse} with minor modifications for compatibility with the version of HOOMD-blue that we used.

\section{Diffusion}
\label{sec:D}
Neutrally buoyant colloids in a suspension move ballistically at times $t \ll \tau_{\rm B}$ before their motion can become diffusive at $t \gtrsim \tau_{\rm B}$. At infinite dilution, the colloid self-diffusion coefficient is $D_0$ for $t \gg \tau_{\rm B}$; however, at finite concentration, the colloids cannot diffuse freely for an indefinite duration. They eventually come into contact with other colloids at times $t \sim \tau_0$, leading to a slowing of their dynamics while their local environment relaxes. At longer times $t \gg \tau_0$, the colloid dynamics can then be viewed as the collective motion of a colloid carrying a ``cage'' of other colloids, resulting in slower dynamics \cite{brady:jfm:1994}. Hence, the diffusive motion of a colloid is commonly described by two diffusion coefficients: a short-time self-diffusion coefficient $D_{\rm S}$ that characterizes local diffusive motion at times $\tau_{\rm B} \lesssim t \lesssim \tau_0$, and a long-time self-diffusion coefficient $D_{\rm L}$ ($\le D_{\rm S}$ for hard spheres) that characterizes the overall relaxation at $t \gg \tau_0$. Note that $D_{\rm S}$ and $D_{\rm L}$ are identical at infinite dilution and equal to $D_0$.

To determine diffusion coefficients in our simulations, we measured the mean squared displacement $\la \Delta r^2 \ra$ of the suspended colloids and computed its time derivative
\begin{equation}
    \alpha(t) = \frac{1}{6}\frac{{\rm d}\la \Delta r^2 \ra}{{\rm d}t} .
    \label{eq:alpha}
\end{equation}
Diffusive motion is characterized by $\la \Delta r^2 \ra \sim 6 D t$ in three dimensions, meaning that $\alpha \sim D$ should be a constant for diffusive motion. The long-time self-diffusion coefficient $D_{\rm L}$ can be determined from a long-time plateau of $\alpha$; we averaged $\alpha$ over $3\,\tau_0 \le t \le 12\,\tau_0$ for our data. The short-time self-diffusion coefficient $D_{\rm S}$ is a little more ambiguous to characterize, and we will discuss this point further in Sec.~\ref{sec:D:short}.

The long-ranged nature of hydrodynamic interactions [see, e.g., Eq.~(\ref{eq:RPY})] can lead to significant finite-size effects in simulations with periodic boundary conditions due to the solvent-mediated interactions of the suspended particles with their periodic images.\cite{beenakker:jcp:1986, duenweg:prl:1991, duenweg:jcp:1993, yeh:jpcb:2004} D{\"u}nweg and Kremer showed that the long-time self-diffusion coefficient $D_{\rm L}^\infty$ for a solute in an infinitely sized simulation box can be obtained from measurements of $D_{\rm L}$ in a cubic simulation box with edge length $L$ using
\begin{equation}
    D_{\rm L}^\infty = D_{\rm L} + \xi \frac{k_{\rm B}T}{6\pi\eta_0 L},
    \label{eq:Dinf}
\end{equation}
where $\xi \approx 2.837297$ plays a role analogous to a Madelung constant.\cite{duenweg:jcp:1993, hasimoto:jfm:1959, yeh:jpcb:2004} Finite-size corrections have also been developed for noncubic periodic\cite{hasimoto:jfm:1959, botan:mp:2015} and partially confined systems\cite{simonnin:jctc:2017} but will not be discussed here.

Equation~\eqref{eq:Dinf} can be readily applied to compute $D_{\rm L}^\infty$ in dilute suspensions ($\phi \to 0$); however, this expression becomes inaccurate even at small volume fractions. In this concentration regime, the functional form of the hydrodynamic interactions remains unchanged,\cite{beenakker:physa:1983} but the interactions must propagate through a medium with an effectively higher viscosity \cite{ladd:jcp:1990, mendoza:jcp:2009, kobayashi:sm:2020}. Hence, finite-concentration effects can be accounted for in Eq.~\eqref{eq:Dinf} by replacing the solvent viscosity $\eta_0$ by the suspension viscosity $\eta$.\cite{ladd:jcp:1990} Determining $\eta$ would require additional simulations, but fortunately, we can approximately eliminate the suspension viscosity from Eq.~\eqref{eq:Dinf} by positing a Stokes--Einstein relationship $D_{\rm L}^\infty = D_0 (\eta_0/\eta)$.\cite{yeh:jpcb:2004}. Substituting for $\eta$ and solving for $D_{\rm L}^\infty$ gives
\begin{equation}
    D_{\rm L}^\infty \approx D_{\rm L}\left(1-\xi \frac{\gamma_0}{6\pi\eta_0 L}\right)^{-1} .
    \label{eq:Dinf2}
\end{equation}
To apply Eq.~\eqref{eq:Dinf2} to our simulations, we assumed that the spherical colloids had hydrodynamic radius $a$ and no-slip boundary conditions in both the MD+MPCD and BD+RPY models, so $\gamma_0 = 6 \pi \eta_0 a$ could be directly substituted. It is important to note that Eqs.~\eqref{eq:Dinf} and \eqref{eq:Dinf2} are based on a low-Reynolds-number description of hydrodynamics and are therefore also subject to the approximations discussed in Sec.~\ref{sec:models:BD}.

To test this approach for correcting finite-size effects, we performed additional MD+MPCD and BD+RPY simulations at $\phi=0.05$ and $\phi=0.30$, varying the box size between $L=20\,a$ and $L=60\,a$, and corrected the measured $D_{\rm L}$ using Eq.~\eqref{eq:Dinf2}. We found that the resulting $D_{\rm L}^\infty$ values at both $\phi$ were independent of $L$ within our measurement accuracy, which provides an a posteriori justification for the assumptions we made. Therefore, we have corrected the long-time self-diffusion coefficients from the MD+MPCD and BD+RPY simulations using Eq.~\eqref{eq:Dinf2} to facilitate the comparison with theory and other simulations. For simplicity, we will not distinguish $D_{\rm L}^\infty$ from $D_{\rm L}$ in the following discussion.

\subsection{Short-time self-diffusion}
\label{sec:D:short}
Within a Brownian dynamics theoretical framework, motion is instantaneously diffusive due to the neglect of colloid inertia, so $D_{\rm S}$ must relate to the displacement of a colloid over a single small timestep $\Delta t$. Accordingly, $D_{\rm S}$ can be determined from the trace of the mobility tensor $\mathcal{M}$ \cite{beenakker:physa:1983},
\begin{equation}
D_{\rm S} = \frac{k_{\rm B} T}{3N} {\rm tr}\,\mathcal{M}.
\label{eq:DSM}
\end{equation}
For pairwise hydrodynamic interactions, it immediately follows that $D_{\rm S} = (k_{\rm B} T/3) {\rm tr}\,\vv{M}^{(ii)}$, so $D_{\rm S} = D_0$ for free-draining hydrodynamics or free-space RPY hydrodynamics [Eq.~\eqref{eq:RPY}], and $D_{\rm S}$ can be evaluated for periodic RPY hydrodynamics using Eq.~\eqref{eq:RPYkspace} \cite{fiore:jcp:2017}. For our simulation box ($L = 40\,a$), we find $D_{\rm S} = 0.929\,D_0$, which should be regarded as a finite-size effect to be corrected. For higher-order hydrodynamic interactions, $D_{\rm S}$ is expected to depend on $\phi$, but determining an analytic expression is challenging. Batchelor derived the first-order correction to the short-time self-diffusion with the two-sphere mobility tensor \cite{batchelor:jfm:1976}, finding $D_{\rm S}(\phi) = D_0(1-1.83\phi)$. Beenakker and Mazur then included second-order correction terms to $D_{\rm S}$ based on approximated near-field hydrodynamic interactions and a partial resummation of the many-body contributions;\cite{beenakker:physa:1984} the predicted $D_{\rm S}$ values were in good agreement with results from experiments and accelerated Stokesian dynamics simulations for $\phi \lesssim 0.4$,\cite{banchio:jcp:2008} despite the lack of lubrication corrections in the theory.

\begin{figure}
    \centering
    \includegraphics{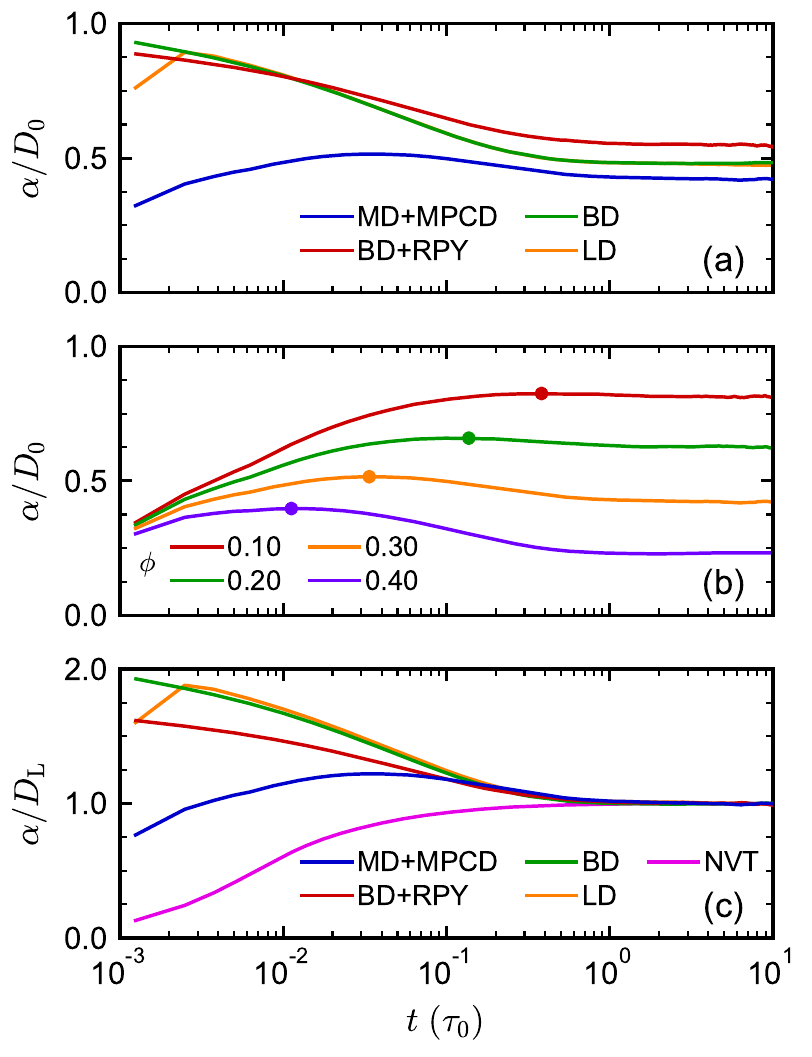}
    \caption{Time derivative $\alpha$ of the colloid mean squared displacement from (a) different methods at $\phi=0.3$ and (b) MD+MPCD at different $\phi$. In (c), $\alpha$ is normalized by $D_{\rm L}$ without finite-size corrections to compare methods in (a) with NVT simulations.}
    \label{fig:alpha}
\end{figure}

Given this theoretical background, we expected both the BD and BD+RPY simulations to give short-time diffusion coefficients $D_{\rm S} \approx D_0$, independent of $\phi$. To approximately extract $D_{\rm S}$ from the simulations, we considered $\alpha(t)$, which should steadily decrease from $D_{\rm S}$ at $t = 0$ to its long-time limit, $D_{\rm L}$. We found that $\alpha$ approximately approached $D_0$ in the BD simulations and a slightly smaller value in the BD+RPY simulations as $t \to 0$ [Fig.~\ref{fig:alpha}(a)]. Both values were consistent with---but slightly smaller than---predicted by Eq.~\eqref{eq:DSM} at $\phi = 0.30$ due to excluded-volume interactions between the colloids during the measurement time, which effectively reduce $\alpha$. We have verified that we exactly obtain theoretically expected values for $D_{\rm S}$ from our simulations at smaller $\phi$ or if we measure the displacement over a single simulation timestep with excluded-volume interactions disabled.

\begin{figure}
    \centering
    \includegraphics{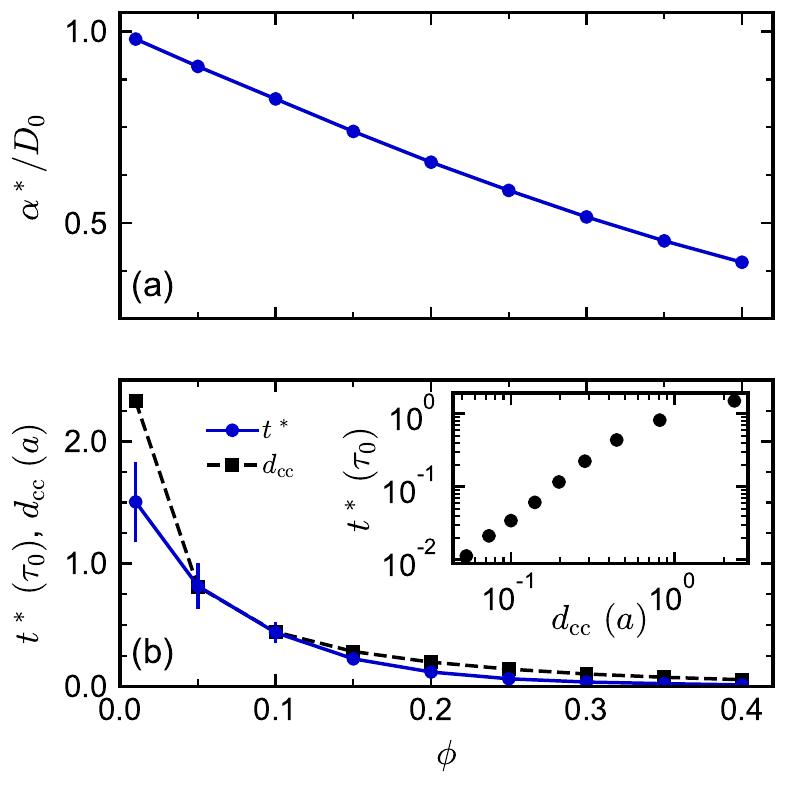}
    \caption{(a) Maximum $\alpha^*$ at time $t^*$ from the MD+MPCD simulations [Fig.~\ref{fig:alpha}(b)]. (b) $t^*$ and theoretical average nearest surface-to-surface distance of colloids $d_{\rm cc}$ as functions of $\phi$. Inset: $t^*$ vs. $d_{\rm cc}$.}
    \label{fig:DSS}
\end{figure}

In stark contrast to the BD and BD+RPY simulations, $\alpha$ was not a monotonic function of $t$ in the MD+MPCD simulations: $\alpha$ increased at short times, obtained a maximum $\alpha^*$ at time $t^*$, then decreased to a long-time plateau for $t \gtrsim \tau_0$ [Fig.~\ref{fig:alpha}(b)]. To rule out the maximum $\alpha^*$ emerging due to inertial effects, which are present in MD+MPCD but not in BD or BD+RPY (Sec.~\ref{sec:models:BD}), we performed LD simulations of single-particle colloids with $\gamma_i = \gamma_0$. We found that $\alpha$ in the LD simulations increased for $t \lesssim 2.5\times 10^{-3}\,\tau_0 \approx 5\,\tau_{\rm B}$, reflecting the ballistic motion of the colloids at short times, but essentially tracked $\alpha$ from the BD simulations after that time  [Fig.~\ref{fig:alpha}(a)]. The maximum $\alpha^*$ in the MD+MPCD simulations occurs, however, at much later times $t^* \gg 5\,\tau_{\rm B}$ [Fig.~\ref{fig:DSS}(b)], and therefore cannot be due to inertia if the colloid experiences a hydrodynamic friction $\gamma_0$.

We considered that the behavior of $\alpha$ in the MD+MPCD simulations might be caused by a lack of coupling between the colloids and the MPCD solvent such that the hydrodynamic friction on a colloid was less than $\gamma_0$ at short times. In the worst-case limit, the colloids would be completely decoupled from the solvent and experience no friction, as in vacuum. To compare the MD+MPCD results with this limit, we performed isothermal--isochoric (NVT) simulations \cite{allen:2017} of the colloids, each represented by a single particle, using a Nos\'{e}--Hoover thermostat with time constant $0.1\,\tau$ and timestep $0.005\,\tau$. The colloid motion was expected to become diffusive in the NVT simulations at long times because of collisions between particles. However, the corresponding long-time self-diffusion coefficient $D_{\rm L}$ was expected to be larger than in the MD+MPCD simulations due to the lack of solvent. Indeed, at $\phi = 0.30$, $\alpha$ was consistently larger in the NVT simulations than in the MD+MPCD simulations, leading to $D_{\rm L}$ being roughly 40 times larger. Accordingly, we normalized $\alpha$ by $D_{\rm L}$ to facilitate comparison with the MD+MPCD simulations [Fig.~\ref{fig:alpha}(c)]. In the NVT simulations, $\alpha/D_{\rm L}$ increased monotonically to one. Its shape at short times is reminiscent of the MD+MPCD simulations at short times, but there was no significant maximum $\alpha^*$ in the NVT simulations. This comparison suggests that there may be \textit{partial} coupling of the colloids to the solvent in MD+MPCD at short times, but more complete hydrodynamic coupling at longer times. This contrasts with the BD and BD+RPY simulations, where the colloids experience an exactly prescribed friction at all times.

The fundamental differences between the MD+MPCD and BD+RPY simulations at short times might be understood by considering the time scales at which hydrodynamic interactions propagate \cite{shakeri:pfl:2018}. In BD+RPY, the solvent-mediated coupling between the colloids is instantaneous and only depends on their positions, whereas in MD+MPCD, shear and sound waves propagate with finite velocities $v_{\rm h} \sim \ell/\nu_0$ and $v_{\rm c} \sim \sqrt{k_{\rm B}T/m} = \ell/\tau$, respectively. The resulting time scales $\tau_{\rm h} \sim a/v_{\rm h}$ and $\tau_{\rm c} \sim a/v_{\rm c}$ are comparable to the typical time $\tau_{\rm B}$ at which Brownian motion of the colloids sets in. It was argued in Ref.~\citenum{shakeri:pfl:2018} that this lack of time scale separation may cause incomplete hydrodynamic coupling at short times.

Nevertheless, it is tempting to identify $\alpha^*$ in the MD+MPCD simulations as the short-time self-diffusion coefficient $D_{\rm S}$ because it decreases with increasing $\phi$ [Fig.~\ref{fig:DSS}(a)] and occurs in the expected time interval $\tau_{\rm B} \lesssim t^* \lesssim \tau_0$ [Fig.~\ref{fig:alpha}(b)]. Previous MD+MPCD simulations that used reflection-based colloid--solvent coupling adopted this definition of $D_{\rm S}$ \cite{bolintineanu:cpm:2014}. We note, however, that this definition is somewhat at odds with the idea that $D_{\rm S}$ should result from hydrodynamic interactions between colloids at times before the colloids physically interact. To interrogate this point, we determined $t^*$ as a function of $\phi$ [Fig.~\ref{fig:DSS}(b)], finding that it decreased as $\phi$ increased. We simultaneously estimated the average surface-to-surface distance $d_{\rm cc}$ between nearest-neighbor colloids using a theoretical expression developed by Torquato for random hard-sphere packings \cite{torquato:pre:1995} [Fig.~\ref{fig:DSS}(b)]. Note that $d_{\rm cc}$ is considerably smaller than the radius of a colloid at most packing fractions and also decreases with $\phi$, implying that two colloids can encounter each other at times much shorter than $\tau_0$. This behavior is consistent with the monotonic decrease in $\alpha$ in the BD and BD+RPY simulations at short times, as colloids should diffuse more slowly when they are hindered or caged by nearby particles.

We found that $t^*$ correlated strongly with $d_{\rm cc}$ [inset of Fig.~\ref{fig:DSS}(b)], suggesting that the maximum in $\alpha$ in the MD+MPCD simulations may be due to interactions between close colloids. However, it is challenging to separate effects from direct interactions [Eq.~\eqref{eq:wca}], from (possible) near-field hydrodynamic interactions, and from lack of separation of timescales in MD+MPCD simulations. Given this ambiguity, we have chosen not to report $\alpha^*$ as $D_{\rm S}$ for the MD+MPCD simulations. The curious short-time behavior in the MD+MPCD simulations should be carefully considered when applying the model to study dynamics. We are, however, encouraged that despite differences at short times, $\alpha$ still has qualitatively the same behavior at intermediate and long times in MD+MPCD as in BD+RPY [Fig.~\ref{fig:alpha}(b)].

\subsection{Long-time self-diffusion}
\begin{figure}
    \centering
    \includegraphics{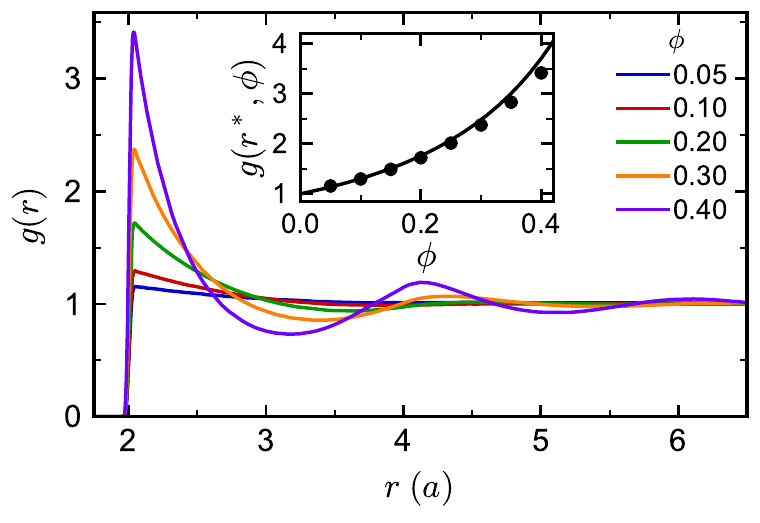}
    \caption{Radial distribution function $g(r)$ between the colloids for selected volume fractions $\phi$. Inset: Value at contact $g(d^+; \phi)$ as a function of $\phi$ from our simulations (symbols) and from Eq.~\eqref{eq:gCS} (solid line).}
    \label{fig:gr}
\end{figure}

The long-time self-diffusion of the colloids is strongly tied to the structure of the suspension. Brady argued that for hard spheres $D_{\rm L}$ can be factored into hydrodynamic and structural contributions, approximately given by\cite{brady:jfm:1994}
\begin{equation}
    D_{\rm L}(\phi) = D_{\rm S}(\phi) \left[1+2\phi g(d^+;\phi)\right]^{-1} ,
    \label{eq:DL}
\end{equation}
where $g(d^+;\phi)$ is the contact value of the radial distribution function $g(r)$ for a given volume fraction $\phi$ as $r \to d$ from the right. Figure~\ref{fig:gr} shows $g(r)$ from our simulations at different $\phi$, demonstrating the well-known behavior that the first peak in $g(r)$ increases with increasing $\phi$ for hard spheres as their local structure becomes more pronounced. For a hard-sphere fluid, the first peak corresponds to $g(d^+;\phi)$, and its value is related to the thermodynamic pressure. We theoretically estimate
\begin{equation}
g(d^+;\phi) = \frac{1-\phi/2}{(1-\phi)^3}
\label{eq:gCS}
\end{equation}
using the Carnahan--Starling equation of state \cite{carnahan:jcp:1969}. We simulated nearly hard spheres [Eq.~\eqref{eq:wca}] rather than true hard spheres, so contact is not as precisely defined for our model: the first peak in $g(r)$ occurs at $r$ slightly larger than $d$. We approximated $g(d^+;\phi)$ in the simulations by the height of the first peak, which we determined by fitting a quadratic interpolating function through the first peak of $g(r)$. The computed $g(d^+;\phi)$ is in excellent agreement with Eq.~\eqref{eq:gCS} (inset of Fig.~\ref{fig:gr}), especially for $\phi \lesssim 0.2$. We accordingly compared $D_{\rm L}$ from our simulations to predictions of Eq.~\eqref{eq:DL} using Eq.~\eqref{eq:gCS} and assuming $D_{\rm S} = D_0$. This value of $D_{\rm S}$ is prescribed in the BD and BD+RPY simulations and is an approximation for the MD+MPCD simulations for lack of well-defined short-time diffusion (Sec.~\ref{sec:D:short}).

\begin{figure}
    \centering
    \includegraphics{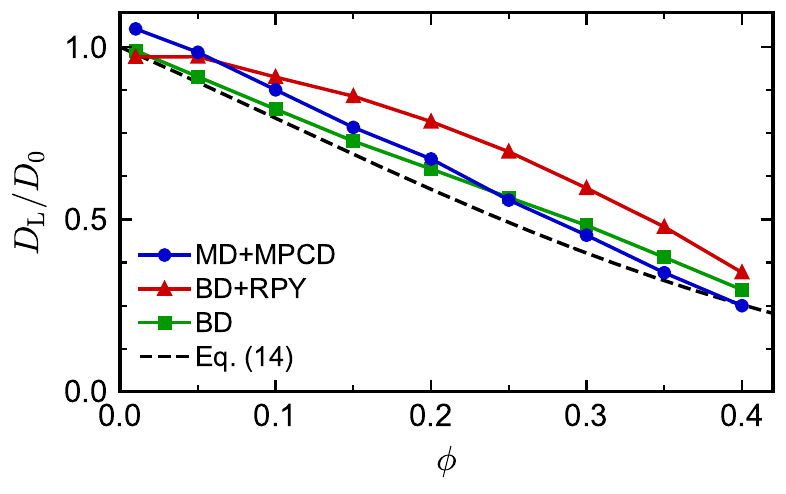}
    \caption{Long-time self-diffusion coefficients $D_{\rm L}$ as functions of volume fraction $\phi$ for the different simulation methods and as predicted by Eq.~\eqref{eq:DL} with Eq.~\eqref{eq:gCS} and $D_{\rm S} = D_0$.}
    \label{fig:DL}
\end{figure}

For small colloid volume fractions $\phi \lesssim 0.1$, the long-time self-diffusion coefficient $D_{\rm L}$ measured in the BD simulations (Fig.~\ref{fig:DL}) was close to the theoretical prediction [Eq.~\eqref{eq:DL}] using $D_{\rm S} = D_0$. At larger $\phi$, however, $D_{\rm L}$ was systematically larger in the BD simulations than theoretically predicted. To test whether this discrepancy originated from the soft colloid--colloid repulsion in our simulations [Eq.~(\ref{eq:wca})], we compared our results with previous Monte Carlo simulations using a true hard-sphere model,\cite{cichocki:bunsen:1990} finding excellent agreement within the measurement accuracy. Further, using the measured $g(d^+;\phi)$ value in Eq.~(\ref{eq:DL}) rather than Eq.~\eqref{eq:gCS} did not significantly change the theoretical prediction. In the BD+RPY simulations, $D_{\rm L}$ was systematically larger than in both the BD simulations and the theoretical prediction. We were surprised to also observe a qualitatively different dependence on $\phi$ given that both the BD and BD+RPY simulations have $D_{\rm S} = D_0$ and give the same suspension structure, so $D_{\rm L}$ should have the same dependence on $\phi$ in both methods according to Eq.~\eqref{eq:DL}. Brady pointed out in his analysis that although Eq.~\eqref{eq:DL} gives the proper asymptotic behavior as $\phi$ approaches maximum packing, it should be regarded as a simple estimate at other $\phi$\cite{brady:jfm:1994}. Evidently, this estimate is inaccurate for BD+RPY at the simulated volume fractions, which are far from maximum packing.

In the MD+MPCD simulations, $D_{\rm L}$ had similar values and dependence on $\phi$ as in the BD simulations and as predicted by Eq.~\eqref{eq:DL} assuming $D_{\rm S} = D_0$. We note that careful inspection of Fig.~\ref{fig:DL} shows that $D_{\rm L}$ was slightly larger (about 5\%) than $D_0$ in the MD+MPCD simulations at $\phi = 0.01$. This discrepancy might be interpreted as a slightly smaller effective friction on the colloid than expected from the Stokes-Einstein relation, which could be explained by an effective hydrodynamic radius that differs from the nominal radius $a$. Following this idea, Poblete et al.~reported the effective hydrodynamic radius for their discrete particle model in MPCD based on single-colloid simulations. The specific value depended on the measurement protocol but was typically larger than $a$ by as much as 10\%. This finding contrasts with our measurement, which would lead to an effective radius about 5\% smaller than $a$ in order to make $D_{\rm L}$ larger than $D_0$ for $\phi = 0.01$. However, we prefer not to attribute this difference to the hydrodynamic size since it is not straightforward to separate the colloid size from other factors (e.g., viscosity) determining the friction coefficient in the MD+MPCD model. A smaller friction coefficient than expected could be related to the extent of colloid--solvent coupling discussed in Sec.~\ref{sec:D:short}.

\section{Sedimentation}
In addition to equilibrium self-diffusion, we also studied nonequilibrium sedimentation of the spherical colloids. Sedimentation is not only an important process in its own right but also helpful for understanding the analogous process of collective diffusion down a concentration gradient.\cite{batchelor:jfm:1976, batchelor:jfm:1983, fortini:prl:2016, howard:lng:2017} During sedimentation, all colloids are subject to a constant external force $\vv{f}$ that causes them to move in the direction of $\vv{f}$ with average velocity $\vv{u}$. If the suspension is incompressible, the motion of the colloids implies an opposing backflow of solvent with average velocity $\vv{u}_0$ such that the volume-average velocity $\vv{u}_{\rm v} = \phi \vv{u} + (1-\phi) \vv{u}_0$ is zero \cite{russel:1989}. Backflow tends to hinder sedimentation of the colloids, and as a result, the rate of sedimentation is experimentally known to depend on the colloid volume fraction $\phi$. This dependence is enhanced by hydrodynamic interactions between colloids.

Sedimentation is conveniently characterized by the sedimentation coefficient $K$, defined by $\vv{u} = K \gamma_0^{-1} \vv{f}$, that relates the average velocity of a colloid in the suspension to the average velocity of a single colloid moving under the same force. Batchelor showed that \cite{batchelor:jfm:1972}
\begin{equation}
    K(\phi) = 1 - 6.55 \phi
    \label{eq:Kbatch}
\end{equation}
for an incompressible hard-sphere suspension at dilute volume fractions using exact two-sphere mobility tensors. Batchelor's result can be extrapolated to higher volume fractions using the empirical expression \cite{russel:1989}
\begin{equation}
    K(\phi) \approx (1-\phi)^{6.55},
    \label{eq:Kemp}
\end{equation}
which has the same small-$\phi$ expansion. Brady and Durlofsky computed the sedimentation coefficient for an incompressible hard-sphere suspension with pairwise far-field hydrodynamics given by the RPY tensor, resulting in
\begin{equation}
    K(\phi) = \frac{(1-\phi)^3}{1+2\phi}
    \label{eq:Krpy}
\end{equation}
after simplification. The small-$\phi$ expansion of this result is $K(\phi) = 1-5\phi$, which is a slightly weaker dependence on $\phi$ than Eq.~\eqref{eq:Kbatch}.

We measured $K$ for our models using nonequilibrium simulations \cite{maginn:jcp:1993} where we applied a constant force $\vv{f} = f_x \vv{\hat{x}}$ to the colloids along the $x$-direction $\vv{\hat{x}}$. In the MD+MPCD simulations, this force was equally distributed to all $N_{\rm v}+1$ constituent particles of a colloid, and an opposing force was applied to the solvent particles to ensure that the entire system was force-free and did not accelerate. (These simulations were performed using double-precision floating-point arithmetic to improve numerical accuracy and momentum conservation \cite{howard:cpc:2018}.) In the BD and BD+RPY simulations, the force was applied to the single particle representing the colloid. 

\begin{figure}
    \centering
    \includegraphics{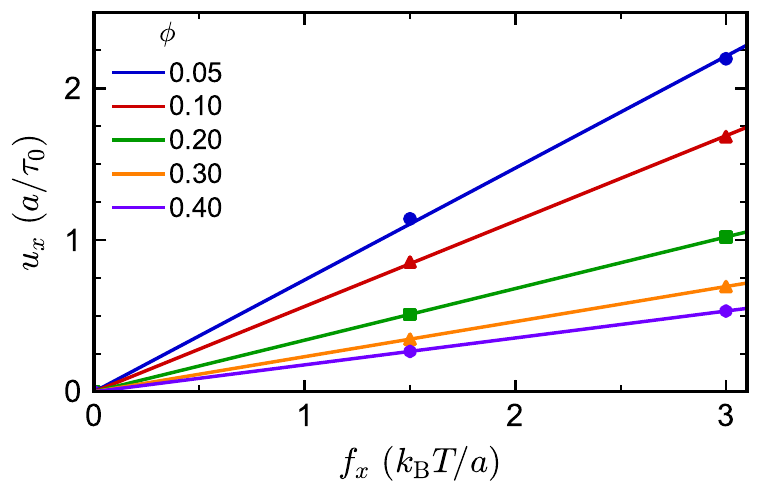}
    \caption{Average colloid velocities $u_x$ in the stationary frame as functions of applied force $f_x$ from the MD+MPCD simulations for all investigated colloid volume fractions $\phi$. The estimated measurement uncertainty is smaller than the symbol size.}
    \label{fig:avgVel}
\end{figure}

For all simulation methods, we applied two forces $f_x = 1.5\,k_{\rm B} T/a$ and $3.0\,k_{\rm B}T/a$ and measured $\vv{u}$ in a stationary frame, that is, the value we observed in the simulations. We simulated $50\,\tau_0$ to reach steady state, then sampled $\vv{u}$ roughly every $5 \times 10^{-5}\,\tau_0$ during a $25\,\tau_0$ simulation. We noted that the colloids in the MD+MPCD simulations accelerated between stochastic collisions because the colloids follow Newton's equations of motion [Eq.~\eqref{eq:MPCDStream}] during this time and experience a net force. We also found that the central and vertex particles oscillated out of phase with each other due to their bonded interactions, which are not damped between collisions. We accordingly computed $\vv{u}$ in the MD+MPCD simulations using the average velocity of all particles comprising a colloid and sampled $\vv{u}$ uniformly over timesteps between collisions to obtain a representative average. In the BD and BD+RPY simulations, we simply computed $\vv{u}$ from the displacement of the colloids during a single time step. We then fit $u_x = \vv{u}\cdot\vv{\hat{x}}$ as a linear function of $f_x$, which is the expected relationship for small $f_x$, and determined $K$ from the slope of the fit. Figure~\ref{fig:avgVel} shows $u_x$ as a function of $f_x$ for the MD+MPCD simulations, and the data are well-fit by a line for all investigated $\phi$.

Similarly to self-diffusion coefficients, sedimentation coefficients measured in simulations with periodic boundary conditions should be corrected for finite-size effects. We used the correction proposed by Ladd \cite{ladd:jcp:1990} for spherical colloids with no-slip boundary conditions to determine the sedimentation coefficient in an infinite-size box $K^\infty$ from measurements of $K$ in a finite-size box. Retaining only the first-order correction, Ladd's result can be rearranged as
\begin{equation}
    K^\infty \approx K \left(1 - \xi \frac{\eta_0 a}{\eta L} \right)^{-1}.
\end{equation}
To eliminate the suspension-viscosity dependence in this correction, we made an assumption consistent with Eq.~\eqref{eq:Dinf2} that $\eta_0/\eta = D_{\rm L}^\infty/D_0$, so
\begin{equation}
    K^\infty \approx K \left(1 - \xi \frac{D_{\rm L}^\infty a}{D_0 L} \right)^{-1}.
\end{equation}
We applied this correction to our measurements, but as for $D_{\rm L}$, we will not retain the notation $K^\infty$ in the discussion that follows.

\begin{figure}
    \centering
    \includegraphics{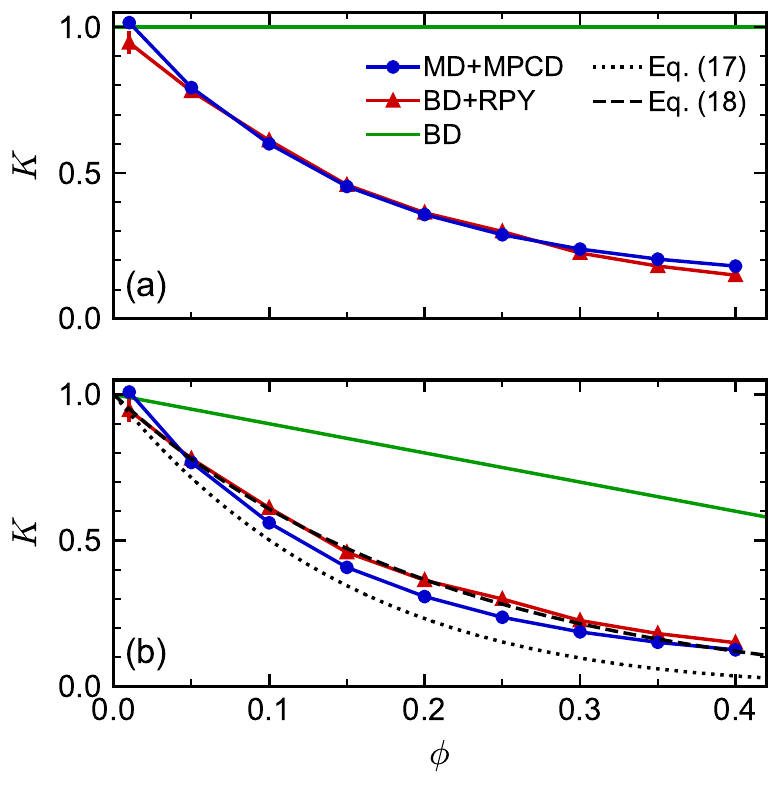}
    \caption{Sedimentation coefficients $K$ as functions of colloid volume fraction $\phi$ for the different simulation methods in (a) the stationary frame observed in the simulations and (b) the frame where the volume-averaged velocity is zero. (Note that the BD+RPY data is the same in both panels.) Theoretical predictions are shown in (b).}
    \label{fig:coeff}
\end{figure}

Figure~\ref{fig:coeff}(a) shows the sedimentation coefficient $K$ as a function of $\phi$ for the MD+MPCD, BD, and BD+RPY simulations. For the BD simulations, $K$ can be analytically determined to be $K = 1$ independent of $\phi$ using a simple argument: The sum of all pairwise forces between colloids must be zero and the time-average of the random displacements is also zero, so averaging the displacements in Eq.~\eqref{eq:BD} over all colloids for BD hydrodynamics leaves only the displacement from the external force $\la \Delta \vv{r} \ra = \la N^{-1} \sum_{i=1}^{N} \Delta \vv{r}_i \ra = \gamma_0^{-1} \vv{f} \Delta t$ and $K=1$. Accordingly, we have plotted only this theoretical value, but we verified that this result was also obtained in BD simulations. In contrast, $K$ decreased strongly as $\phi$ increased in the BD+RPY and MD+MPCD simulations and was nearly the same for both methods at all $\phi$ investigated. As expected, $K \to 1$ as $\phi \to 0$ in the BD+RPY simulations, but in the MD+MPCD simulations, $K$ was slightly larger than one at $\phi = 0.01$. This discrepancy is consistent with $D_{\rm L}$ being somewhat larger than $D_0$ at the same concentration (Fig.~\ref{fig:DL}).

In order to compare these results with each other and with theoretical predictions, it is important to note that $K$ depends on the reference frame chosen to measure $\vv{u}$ \cite{brady:ajs:1975, howard:jcp:2020}. In both Batchelor's and Brady and Durlofsky's analyses, the sedimentation velocities are defined relative to $\vv{u}_{\rm v}$, which is zero for an incompressible suspension because there is no volume flux. The MD+MPCD simulations, however, were effectively performed in a reference frame where the mass-averaged velocity $\vv{u}_{\rm m} = w \vv{u} + (1-w) \vv{u}_0$, with $w$ being the colloid mass fraction, is zero due to conservation of linear momentum. Having $\vv{u}_{\rm m} = \vv{0}$ does not also give $\vv{u}_{\rm v} = \vv{0}$ for our MD+MPCD model, so the colloid velocities were effectively measured relative to $\vv{u}_{\rm m}$ rather than $\vv{u}_{\rm v}$. The BD equations of motion imply a stationary solvent, and so the colloid velocities were measured relative to the solvent velocity rather than $\vv{u}_{\rm v}$. The BD+RPY simulations have a backflow of solvent that results from discarding the $\vv{k} = \vv{0}$ wavevector in Eq.~\eqref{eq:RPYkspace} such that $\vv{u}_{\rm v} = \vv{0}$ \cite{brady:jfm:1988}, so these simulations have the same reference frame as the theoretical calculations.

Accordingly, we shifted $\vv{u}$ and $\vv{u}_0$ measured in the MD+MPCD and BD simulations to $\vv{u}' = \vv{u} - \vv{u}_{\rm v}$ and $\vv{u}_0' = \vv{u}_0 - \vv{u}_{\rm v}$ so that $\vv{u}_{\rm v}' = \vv{0}$ in the shifted reference frame. For the MD+MPCD simulations, $\vv{u}' = (1-\phi)[1+w/(1-w)]\vv{u}$, with $w = M N/(M N + m \rho_0 L^3)$, because $\vv{u}_0 = -[w/(1-w)] \vv{u}$ in the frame where $\vv{u}_{\rm m} = \vv{0}$. In performing this shift, we note that there is some ambiguity in defining the colloid and solvent volume fractions in the MD+MPCD model because the colloids do not exclude volume to the solvent; we used the nominal values $\phi$ and $1-\phi$ for consistency with the other simulations. For the BD simulations, $\vv{u}' = (1-\phi) \vv{u}$ because $\vv{u}_0 = \vv{0}$ in the frame that moves with the solvent. These scaling factors between $\vv{u}$ and $\vv{u}'$ can be straightforwardly applied to transform the measured $K$.

After applying the shift to a common reference frame [Fig.~\ref{fig:coeff}(b)], $K$ remained consistently smaller in the MD+MPCD and BD+RPY simulations than in the BD simulations due to the presence of hydrodynamic interactions between colloids. As expected, $K$ for the BD+RPY simulations was in nearly perfect agreement with Eq.~\eqref{eq:Krpy}. The values of $K$ in the MD+MPCD simulations were typically larger than Eq.~\eqref{eq:Kemp} but smaller than Eq.~\eqref{eq:Krpy}. At small $\phi$, the concentration dependence of $K$ was qualitatively similar to Eq.~\eqref{eq:Kemp} but was quantitatively larger; at larger $\phi$, $K$ more closely resembled the results of the BD+RPY simulations.

Overall, the MD+MPCD simulations showed sedimentation behavior that was consistent with the presence of hydrodynamic interactions between colloids. This observation is somewhat at odds with a recent report\cite{shakeri:pfl:2018} that found essentially no long-ranged hydrodynamic interactions between two mutually diffusing colloids modeled using MD+MPCD with a reflection scheme for the solvent--colloid coupling; their conclusion was that MD+MPCD was not suitable for modeling colloidal suspensions. Given the fundamentally different nature of their measurements and ours, it is challenging to make a conclusive comparison but this discrepancy surely deserves further consideration.

\section{Diffusion of cubic colloids}
Having shown that MD+MPCD with a discrete particle model largely reproduces the theoretically expected dynamics of spherical colloids, we extended the model to cubic colloids to demonstrate the generality and flexibility of the approach. The diameter (edge length) of the cubes was set to $d=6\,\ell$, and vertex particles were placed on their surface in a square lattice with spacing $6/7\,\ell$, i.e., 7 vertex particles per edge and $N_{\rm v} = 296$ vertex particles per colloid [Fig.~\ref{fig:model}(b)]. Each vertex particle was connected to its nearest neighbor and its diametrically opposed vertex particle using a harmonic potential [Eq.~(\ref{eq:Ub})]. Unlike the spherical colloids, we did not include a central particle to model the excluded-volume interactions between cubes because these interactions do not have a simple mathematical form. Instead, all nonbonded vertex particles interacted with each other through the Weeks--Chandler--Andersen pair potential \cite{weeks:jcp:1971} [Eq.~\eqref{eq:wca} with $\sigma = 1.0\,\ell$ and $\Delta = 0.0\,\ell$]. We confirmed that the vertex-particle spacing was small enough that the cubes did not penetrate each other. As a side effect of this model for the interactions, each cube had a slightly corrugated texture and excluded a larger volume $v_{\rm ex} \approx (d+\sigma)^3$ than expected from its hydrodynamic diameter, $v = d^3$. In dilute suspensions where colloid--colloid interactions are predominantly solvent-mediated, the hydrodynamic volume fraction $\phi = Nv/L^3$ may dictate behavior, whereas in concentrated suspensions where excluded volume interactions dominate, the thermodynamic volume fraction $\phi_{\rm ex} = Nv_{\rm ex}/L^3$ ($\approx 1.6\,\phi$ for the chosen parameters) may be more relevant. We restricted our simulations to $\phi_{\rm ex} < 0.40$ to stay well below the freezing transition of hard cubes.\cite{agarwal:nm:2011} The size of the simulation box was set to $L=120\,\ell$, and we performed four independent simulations per volume fraction using the same MPCD parameters as for the spherical colloids (Sec.~\ref{sec:models:solvent}).

\begin{figure}[htbp]
    \centering
    \includegraphics{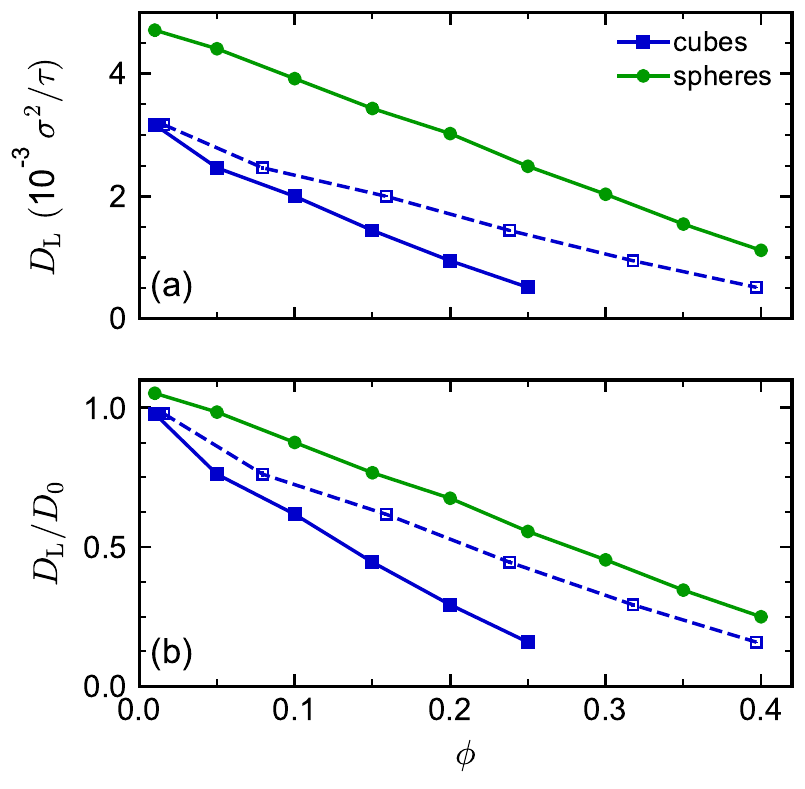}
    \caption{Long-time self-diffusion coefficients $D_{\rm L}$ of cubic and spherical colloids as functions of their volume fraction $\phi$. Filled and open symbols indicate definition of $\phi$ using the hydrodynamic size ($v$) and the excluded volume ($v_{\rm ex}$) of the cubes, respectively. Results shown (a) in MD+MPCD units, and (b) reduced by the theoretically expected self-diffusion coefficient at infinite dilution, $D_0$.}
    \label{fig:cubes}
\end{figure}

Okada and Satoh numerically determined the translational diffusion coefficient of a cube at infinite dilution by measuring the drag force under uniform low-Reynolds-number flow.\cite{okada:mp:2020} They found that the translational friction coefficient $\gamma_0$ was independent of orientation \cite{brenner:1966}. The resulting colloid diffusion coefficient $D_0$ was reduced by a factor of $1.384$ compared to a sphere with the same diameter. Interestingly, this factor is comparable to the mean diameter of the spheres that inscribe and circumscribe the cube, $(1+\sqrt{3})/2\,d \approx 1.366\,d$. Thus, cubes are predicted to diffuse much slower than spheres with the same diameter $d$. Even if one adjusts the cube diameter to match the sphere volume, the resulting $D_0$ of a cube would be about $10\,\%$ smaller than that of a sphere. In the simulations, we can determine $D_0$ from $D_{\rm L}$ in the dilute regime $\phi \to 0$. We applied the same finite size correction [Eq.~\eqref{eq:Dinf2}] as for the spheres, but using a friction coefficient of $\gamma_0 \approx 13.04 \eta_0 d$ for the cubes. Figure~\ref{fig:cubes} show the adjusted $D_{\rm L}$ as functions of $\phi$ and $\phi_{\rm ex}$. At $\phi=0.01$, we find $D_{\rm L} = (3.17 \pm 0.06) \times 10^{-3}\,\ell^2/\tau$, which is close to the theoretically expected value of $D_0 = 3.23 \times 10^{-3}\,\ell^2/\tau$. The diffusivity decreases with increasing colloid volume fraction, as expected, and $D_{\rm L}$ approaches zero as $\phi_{\rm ex}$ comes close to the freezing transition.\cite{agarwal:nm:2011}

\section{Conclusions}
We studied the diffusion and sedimentation of colloidal particles suspended in a Newtonian solvent using various computer simulation techniques. We focused in particular on a discrete particle model for the colloids coupled to the multiparticle collision dynamics model for the solvent (MD+MPCD). We characterized the short-time and long-time motion of spherical colloids, and determined the long-time self-diffusion coefficient and sedimentation coefficient, spanning a range of colloid volume fractions from $\phi=0.01$ to $\phi=0.40$. We compared the MD+MPCD model to theoretical predictions based on low-Reynolds-number hydrodynamics and additional Brownian dynamics simulations using both free-draining hydrodynamics (BD) and pairwise far-field hydrodynamics (BD+RPY).

Overall, we found that the MD+MPCD model for spherical colloids gave satisfactory results that were generally similar to the theoretical predictions and BD+RPY simulations. The largest discrepancy was found for the short-time motion of the colloids: a maximum was observed in the time-derivative $\alpha$ of the colloid mean squared displacement in the MD+MPCD model that was absent from the BD and BD+RPY models. As $\phi$ increased, this maximum decreased and occurred at earlier times; we found that this time correlated with the average distance between nearest colloids. By careful comparison with additional simulations, we posited that this maximum should not be identified as a short-time self-diffusion coefficient, but rather might be due to partial hydrodynamic coupling of the colloids to the solvent in MD+MPCD at short times. However, at longer times, the MD+MPCD model showed many expected features for colloidal suspensions, including a decrease of $\alpha$ to a long-time plateau that gave a long-time self-diffusion coefficient that depended on $\phi$ and was largely consistent with the other models. Further, the measured sedimentation coefficients were almost identical in the MD+MPCD and BD+RPY simulations in the stationary frame; both differed from BD simulations with free-draining hydrodynamics, suggesting the presence of hydrodynamic interactions in the MD+MPCD simulations.

One major benefit of the discrete particle model we used is its extensibility to particle shapes other than spheres. To demonstrate this flexibility, we also studied the long-time self-diffusion of cubic colloidal particles using MD+MPCD. Compared to a spherical colloid of the same diameter, the long-time self-diffusion coefficient of a cube at nearly infinite dilution ($\phi = 0.01$) was reduced by a factor of roughly 1.5, in reasonable agreement with theoretical predictions \cite{okada:mp:2020}. At all investigated colloid volume fractions $\phi$, the cubes diffused slower than spheres of equal diameter. Altogether, we found MD+MPCD to be a convenient and reasonable approach to study colloidal suspensions, especially for particles with complex shapes.

\begin{acknowledgements}
We thank Zachary Sherman for many helpful discussions on this topic. This work was completed in part with resources provided by the Auburn University Easley Cluster. Computing time was also granted on the supercomputer Mogon at Johannes Gutenberg University Mainz (www.hpc.uni-mainz.de). This work was funded by the Deutsche Forschungsgemeinschaft (DFG, German Research Foundation) through projects 274340645; 405552959; NI 1487/7-1.
\end{acknowledgements}

\section*{Data Availability}
The data that support the findings of this study are available from the authors upon reasonable request.

\bibliography{references}

\end{document}